\documentclass{article}
\usepackage{spconf,amsmath,graphicx,amssymb,float,cuted,capt-of}
\usepackage{placeins}
\usepackage{booktabs}
\usepackage{xcolor}
\usepackage[bottom]{footmisc}
\usepackage[hidelinks]{hyperref}

\title{Behavior-Conditioned Animal Vocalization Generation: Event-Aware Modeling and Multidimensional Evaluation}

\name{
Xinlong Deng$^{1}$\sthanks{Equal contribution.},
Yong Tian$^{1*}$,
Jian Guan$^{2}$,
Qiuqiang Kong$^{3}$,
Jie Jiang$^{1}$,
Yin Cao$^{4}$
}

\address{
$^{1}$ College of Artificial Intelligence, China University of Petroleum, Beijing \\
$^{2}$College of Computer Science and Technology, 
Harbin Engineering University, Harbin, China \\
$^{3}$ Department of Electronic Engineering, The Chinese University of Hong Kong, Hong Kong, China \\
$^{4}$ Institute of Acoustics, Chinese Academy of Sciences
}

\begin{document}
\ninept
\maketitle

\begin{abstract}
Behavior-conditioned animal vocalization generation aims to synthesize vocalizations from natural-language descriptions of animal behaviors, social interactions, and contexts. This task remains largely unexplored, with limited behavioral text supervision and substantial variation in vocalization duration. We construct the Behavior-Vocalization Dataset (BeVo) by integrating public bioacoustic data from six species with natural-language behavioral conditions derived from species-specific ethological and bioacoustic literature, totaling 31,191 events. We then adapt a rectified-flow-based generator with Event-Aware variable-length modeling, which encodes vocalizations at their actual durations and uses masks to prevent attention to padded latent positions and exclude these positions from the training loss. We evaluate vocalization discriminability, acoustic distribution fidelity, and language-aligned representation fidelity. Experiments show that task-specific adaptation improves all evaluated metrics compared with its zero-shot baseline, while Event-Aware modeling further increases vocalization-anchor macro accuracy from 59.86\% to 61.99\% and reduces Fréchet Audio Distance from 1.40 to 1.22. Conditioning analyses further show that behavioral conditions improve target-anchor discriminability, while swapping behavior shifts generations toward the corresponding vocalization anchors. Demos are available on our project page.\footnote{\url{https://persimmontian.github.io/BCAVG}}

\end{abstract}

\begin{keywords}
animal vocalization generation, behavior-conditioned generation, bioacoustics, text-to-audio
\end{keywords}

\section{Introduction}
\label{sec:introduction}
Animal vocalizations convey information about behavioral states, social interactions, and communicative contexts, making them an important medium for studying animal communication~\cite{seyfarth2010production,briefer2012vocal}. While species or call-type labels specify which animal or type of vocalization to generate, behavioral descriptions specify the behavioral and social contexts associated with that vocalization. Modern Text-to-Audio (TTA) models, from diffusion-based AudioLDM 2~\cite{liu2024audioldm2} to flow-matching-based TangoFlux~\cite{hung2026tangoflux}, have enabled audio generation from natural-language descriptions. These advances provide a basis for generating animal vocalizations conditioned on natural-language descriptions of behaviors and social contexts.

Animal vocalization generation has evolved from conventional class-conditioned synthesis for data augmentation toward more flexible generative modeling. ORCA-WHISPER~\cite{bergler2022orcawhisper} and MAE-GAN~\cite{wang2023maegan} explored GAN-based animal vocalization generation for low-resource killer whale calls and few-shot dolphin whistles. ECOGEN~\cite{guei2024ecogen} further generated bird vocalizations to augment under-represented classes in ecological sound recognition. GmSLM~\cite{sternberg2025gmslm} introduced generative spoken-language modeling for marmoset vocal communication, extending generation from individual vocalizations to longer sequences and conversational structure. More closely related to our task, BirdDiff~\cite{song2026birddiff} explored controllable diffusion-based bird vocalization generation using category labels, acoustic features, and fixed species-level textual descriptions.

However, existing animal vocalization generation still relies mainly on species, call types, or species-level text, leaving behavior-conditioned generation largely unexplored. BioLingual~\cite{robinson2024biolingual} and NatureLM-audio~\cite{robinson2025naturelm} demonstrate the potential of language-supervised representations and audio–language modeling for bioacoustic understanding. However, existing bioacoustic corpora typically provide behavior-related information through discrete annotations, while paired natural-language descriptions suitable for behavior-conditioned generation remain limited. This limits their direct use as supervision for behavior-conditioned generation. Animal vocalizations are also typically short and duration-variable, causing fixed-length training to introduce substantial padding-derived latent positions that do not correspond to the vocalization event.

\begin{figure*}[t]
    \centering
    \includegraphics[width=\textwidth]{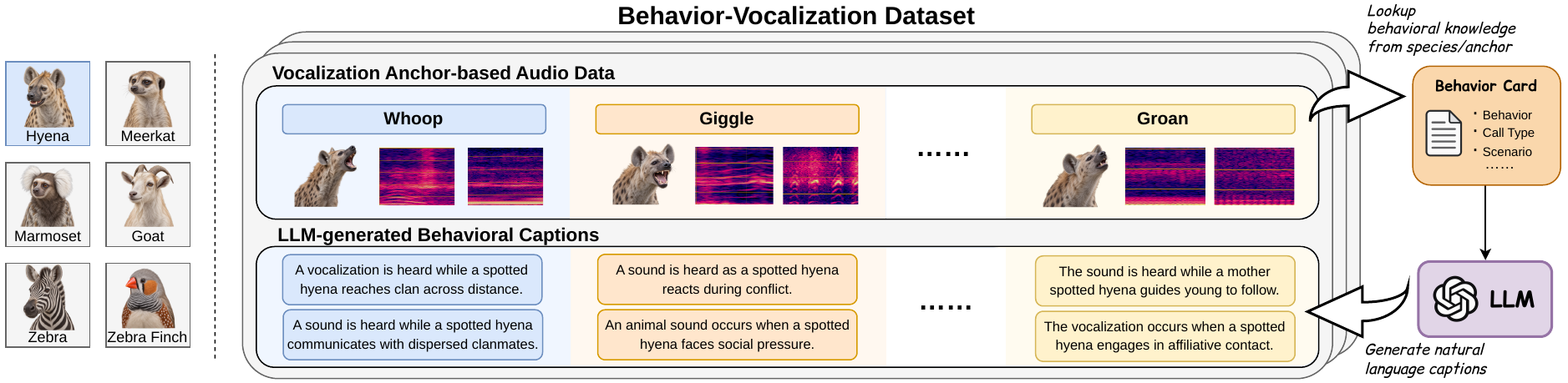}
    \vspace{-4mm}
    \caption{Overview of BeVo dataset construction. Recordings from six species are organized by vocalization anchors and linked to literature-grounded behavior cards. An LLM uses these cards to generate natural-language behavioral descriptions, which are paired with recordings sharing the corresponding anchor.}
    \label{fig:dataset-overview}
\end{figure*}

To address these challenges, we investigate behavior-conditioned animal vocalization generation with the following contributions:
\begin{itemize}
    \setlength{\itemsep}{0pt}
    \setlength{\parsep}{0pt}
    \setlength{\topsep}{0pt}
    \setlength{\partopsep}{0pt}
    
    \item We construct the literature-grounded Behavior-Vocalization Dataset (BeVo) by linking public bioacoustic recordings from six species with descriptions of behaviors, social interactions, and contexts through their existing annotations.

    \item Building on BeVo, we adapt TangoFlux to natural-language behavioral conditions and introduce Event-Aware variable-length modeling, which encodes vocalizations at their actual durations and excludes padded latent positions from attention and rectified-flow optimization.

    \item We develop a multidimensional evaluation framework that assesses vocalization discriminability and distribution fidelity in both acoustic and language-aligned representation spaces, including Acoustic Feature Distribution Distance (AFDD) for interpretable acoustic comparisons.
\end{itemize}



\section{The BeVo Dataset}
\label{sec:dataset}

Public bioacoustic datasets mainly provide discrete species, call-type, behavioral, or contextual annotations, but systematic natural-language supervision for behavior-conditioned generation remains limited. We construct \textbf{BeVo}, covering six species, \textbf{31,191 vocalization events}, and \textbf{42 vocalization anchors} linking recordings to literature-grounded behavioral descriptions, totaling approximately \textbf{524.3 min} of audio. The BeVo dataset is publicly available on this page.\footnotemark[2]Fig.~\ref{fig:dataset-overview} summarizes the dataset construction.

\vspace{-3mm}
\subsection{Audio Data Collection and Curation}
\label{subsec:audio-curation}

We aggregate six public bioacoustic corpora covering spotted hyena (HyenaSET)~\cite{woerner2026hyenaset}, meerkat (MeerKAT)~\cite{schaeferzimmermann2026animal2vec}, common marmoset (MarmAudio)~\cite{lamothe2025marmaudio}, goat (Goat Bleatings)~\cite{ntalampiras2023goat}, plains zebra~\cite{xie2024zebra}, and zebra finch~\cite{elie2016zebrafinch}. Because these corpora differ in recording organization and annotation granularity, we standardize them into \emph{vocalization events}, which we define operationally as temporally localized, annotation-bearing audio units treated as individual samples. HyenaSET and MeerKAT are segmented using annotated temporal boundaries, while the other corpora provide pre-segmented or short vocalization recordings used directly as sample units.

To further improve data quality, we apply additional curation to the collected recordings. To remove low-energy and acoustically anomalous samples, global RMS and local frame-level RMS are examined using thresholds defined separately for different species and vocalization anchors. For suitable corpora, we further perform zero-shot filtering with AnimalCLAP~\cite{shinoda2026animalclap} or BioLingual~\cite{robinson2024biolingual} to remove recordings with low audio--text similarity to the target species. Finally, larger corpora are downsampled to reduce imbalance across species and vocalization anchors. Detailed preprocessing settings and thresholds will be provided on the project page.

\vspace{-3mm}
\subsection{Training and Evaluation Set Construction}
\label{subsec:dataset-split}

\begin{table}[!b]
    \centering
    \caption{Number of vocalization events in each data split.}
    \label{tab:dataset-splits}
    \scriptsize
    \setlength{\tabcolsep}{6pt}
    \begin{tabular}{cccccccc}
        \hline
        Split & Goat & Hyena & Marmo. & Meerkat & Zebra & Z. Finch & Total \\
        \hline
        Train & 2826 & 10216 & 4800 & 4475 & 353 & 2298 & 24968 \\
        Val.  & 360  & 1226  & 600  & 564  & 62  & 311  & 3123 \\
        Test  & 400  & 800   & 600  & 600  & 200 & 500  & 3100 \\
        \hline
        Total & 3586 & 12242 & 6000 & 5639 & 615 & 3109 & 31191 \\
        \hline
    \end{tabular}
\end{table}

To reduce duration-distribution shifts across data splits, samples within each species and vocalization anchor are divided into five duration-quantile strata. We first construct the test set by stratified random sampling, with 100 samples per anchor for spotted hyena, common marmoset, and meerkat, and 50 for goat, plains zebra, and zebra finch. The validation size is set to 10\% of the original samples in each anchor and is then stratified-sampled from the remaining samples using the same duration strata. All remaining samples form the training set, with the final split statistics reported in Table~\ref{tab:dataset-splits}.

\subsection{Literature-Grounded Behavioral Conditions}
\label{subsec:behavior-captions}

\footnotetext[2]{\url{https://huggingface.co/datasets/Xinghour/BeVo}}

For each vocalization anchor, we review species-specific ethological and bioacoustic literature and construct a behavior card summarizing literature-supported behaviors, interaction roles, and contexts. The supporting references are retained in each behavior card.

With these cards, we use GPT-5.6 to generate 500 distinct conditions per anchor. The prompt treats the behavior card as the primary behavioral source, prohibits unsupported inferences and anchor-identifying labels, and applies structural and diversity constraints with self-checking and regeneration; no manual post-generation verification is performed. We hold out 10\% of these conditions for validation, while training and validation events are randomly paired with conditions from their respective anchor-specific pools. Because the available source corpora generally lack event-level behavioral annotations, the conditions are associated with groups of recordings sharing the same vocalization anchor rather than serving as one-to-one behavioral descriptions of individual recordings.

For testing, GPT-5.6 generates 100 conditions per anchor while being provided with previously generated conditions and instructed to avoid repetition. Test events are assigned fixed conditions from this pool. This setup evaluates generalization to new linguistic formulations of behavioral knowledge represented during training, rather than to previously unseen behaviors. We further derive behavior-only and species-only conditions for conditioning analysis.


\section{Behavior-Conditioned Generation with Event-Aware Variable-Length Modeling}
\label{sec:model}

\subsection{Behavior-Conditioned Rectified Flow}
\label{subsec:flow-matching}

We adopt TangoFlux~\cite{hung2026tangoflux} as the generative backbone for behavior-conditioned animal vocalization generation. TangoFlux performs rectified flow matching in the latent space of a pretrained audio autoencoder, conditioned on FLAN-T5 text representations and target duration. Starting from its pretrained checkpoint, we fine-tune the model using the natural-language behavioral conditions constructed in Sec.~\ref{subsec:behavior-captions}.

Following TangoFlux, the conditional velocity field is optimized with the rectified-flow objective
\begin{equation}
\mathcal{L}_{\mathrm{RF}}
=
\mathbb{E}
\left[
\left\|
\mathbf{v}_{\theta}(\mathbf{z}_t,t\mid \mathbf{c},d)
-
(\boldsymbol{\epsilon}-\mathbf{z}_0)
\right\|_2^2
\right],
\label{eq:rectified-flow}
\end{equation}
Here, $\mathbf{z}_0$ is the target audio latent, $\boldsymbol{\epsilon}$ is Gaussian noise, $\mathbf{z}_t$ is obtained by linearly interpolating between the target audio latent and Gaussian noise, and $\mathbf{v}_{\theta}$ is the velocity field parameterized by $\theta$; $\mathbf{c}$ and $d$ denote the text and target-duration conditions. During training, $\mathbf{c}$ is resampled each epoch as species-and-behavior, behavior-only, species-only, or empty text with probabilities $0.80$, $0.10$, $0.05$, and $0.05$, respectively, while $d$ is retained in all cases.

\vspace{-3mm}
\subsection{Event-Aware Variable-Length Modeling}
\label{subsec:variable-length}
\vspace{-1mm}

Animal vocalizations are typically short acoustic events and exhibit substantial variation in duration. Although TangoFlux supports duration conditioning, it uses fixed-length waveform encoding during training. When applied to short animal vocalizations, this fixed-length procedure introduces padding-derived latent positions that do not correspond to the actual vocalization. As a result, a substantial portion of the sequence and training objective may be devoted to non-vocalization regions rather than the acoustic event itself.

\begin{figure}[H]
    \centering
    \includegraphics[width=\columnwidth]{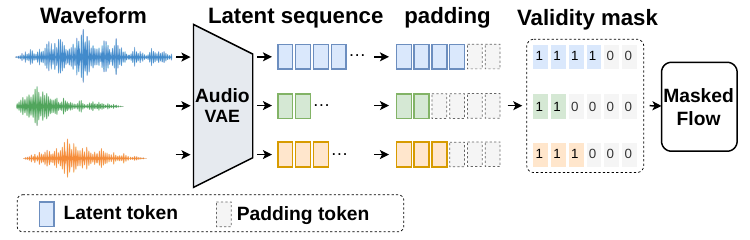}
    \vspace{-4mm}
    \caption{Overview of the event-aware latent modeling.}
    \label{fig:event-aware-variable-length}
\end{figure}

To address this issue, we introduce Event-Aware Variable-Length Modeling, as illustrated in Fig.~\ref{fig:event-aware-variable-length}. Each vocalization is independently encoded according to its actual duration, with only the minimum waveform padding required for VAE hop alignment, thereby avoiding fixed-length padding before VAE encoding. Latent sequences of different lengths are then padded only within each minibatch, and a binary mask is constructed according to the valid latent length of each sample.

\vspace{-3mm}

The mask is applied to both the Flux Transformer attention and the rectified-flow objective, such that padded latent positions neither serve as valid context nor contribute to optimization. For sample $i$, the masked objective is defined as
\begin{equation}
\mathcal{L}_{\mathrm{RF}}^{(i)}
=
\frac{
\sum_j m_{i,j}
\left\|
v_{\theta}(z_{t,i,j}, t \mid c_i,d_i)
-
(\epsilon_{i,j}-z_{0,i,j})
\right\|_2^2
}{
\sum_j m_{i,j}
},
\label{eq:masked-rf}
\end{equation}
where $i$ indexes samples, $j$ indexes latent positions, and $m_{i,j}\in\{0,1\}$ indicates whether position $j$ is valid for sample $i$.

During inference, the latent sequence is initialized according to the requested target duration rather than a predefined maximum length, avoiding unnecessary generation beyond the target event duration. These modifications are evaluated jointly as the Event-Aware modeling scheme.

\vspace{-2mm}
\begin{table*}[t]
    \centering
    \caption{Overall results on behavior-conditioned animal vocalization generation.}
    \label{tab:main-results}
    \setlength{\tabcolsep}{6pt}
    \begin{tabular}{lcccccc}
        \hline
        System
        & Macro Acc. $\uparrow$
        & mAP $\uparrow$
        & FAD $\downarrow$
        & AFDD $\downarrow$
        & AnimalCLAP-FAD $\downarrow$
        & BioLingual-FAD $\downarrow$ \\
        \hline

        \textcolor{gray}{Real Reference}
        & \textcolor{gray}{0.8852}
        & \textcolor{gray}{0.9240}
        & \textcolor{gray}{0.2321}
        & \textcolor{gray}{0.2266}
        & \textcolor{gray}{0.0075}
        & \textcolor{gray}{0.0275} \\

        \hline

        TangoFlux Zero-shot
        & 0.1787 & 0.2022 & 6.7080 & 3.4759 & 0.3318 & 0.4904 \\

        TangoFlux Fine-tuned
        & 0.5986 & 0.6345 & 1.4047 & 1.2480 & \textbf{0.0491} & 0.1237 \\

        TangoFlux Event-Aware
        & \textbf{0.6199} & \textbf{0.6551} & \textbf{1.2239} & \textbf{0.9206} & 0.0492 & \textbf{0.1231} \\

        \hline
    \end{tabular}
    \vspace{-4mm}
\end{table*}


\section{Multidimensional Evaluation Framework}
\label{sec:benchmark}

A single metric is insufficient to characterize all aspects of generation quality. We therefore evaluate target-vocalization discriminability, acoustic distribution fidelity, and language-aligned representation fidelity. Class-conditioned scores are averaged equally over anchors within each species and then across species, preventing more frequent anchors or species from dominating the aggregate scores.

\subsection{Vocalization Discriminability}
\label{subsec:recognizability}
\vspace{-1mm}

Generated vocalizations should retain the acoustic characteristics of the target vocalization anchor. We fine-tune an AVEX/BEATs evaluator~\cite{miron2026avex} with attention pooling and species-specific linear anchor heads using only real training audio: the pooling and heads are optimized first, followed by joint backbone fine-tuning. The checkpoint is selected on real validation audio and frozen for real-test and generated-audio evaluation. Each generated clip inherits its anchor from the behavior card used to construct its prompt. We report Vocalization-Anchor Macro Accuracy for hard decisions and mAP for class-ranking quality; both quantify anchor-level conditional response rather than direct behavioral understanding.

\vspace{-3mm}
\subsection{Acoustic Distribution Fidelity}
\label{subsec:acoustic-fidelity}
\vspace{-1mm}

Discriminability alone does not ensure acoustic realism. FAD~\cite{kilgour2019fad} measures real--generated distribution shift in frozen VGGish~\cite{hershey2017cnn} space; it is computed separately for each species using its complete real-test set and corresponding generated set, then macro-averaged across species. To expose interpretable discrepancies, we further use Acoustic Feature Distribution Distance (AFDD) over 13 descriptors grouped as temporal (active-event duration, temporal centroid, envelope modulation rate/depth), frequency-location (dominant frequency, spectral median/centroid), spectral-shape (energy-IQR bandwidth, entropy, flatness, slope), and spectro-temporal (spectral flux, dominant-frequency excursion). For species $s$, anchor $a$, and descriptor $k$,
\begin{equation}
 d_{s,a,k}=\frac{W_1(\widehat P^r_{s,a,k},\widehat P^g_{s,a,k})}
 {\max\{\mathrm{IQR}^r_{s,a,k},\,0.1\,\mathrm{IQR}^r_{s,\cdot,k},\,\epsilon_k\}}.
\label{eq:afdd}
\end{equation}
Descriptors are extracted from detected active events; spectral analysis uses a 20-ms Hann window with a 5-ms hop, and multiple events in one recording are summarized by their median. Duration, modulation-rate, and frequency-scale descriptors are log-transformed before comparison. Fixed descriptor-specific floors $\epsilon_k$ are applied after transformation solely as numerical safeguards for near-degenerate real distributions.\footnote{Floors by group are: temporal $(0.02,0.01,0.05,0.01)$; frequency-location $(0.02,0.02,0.02)$; spectral-shape $(0.05,0.01,0.01,0.25)$; and spectro-temporal $(0.01,0.02)$.} AFDD first averages descriptors within each acoustic group, assigns equal weight to the four groups, and then macro-averages over anchors and species, preventing descriptor-rich groups from dominating the score. FAD therefore captures global distribution shift, whereas AFDD provides attribute-level diagnosis.

\vspace{-3mm}
\subsection{Language-Aligned Representation Fidelity}
\label{subsec:semantic-fidelity}
\vspace{-1mm}

Acoustic agreement does not necessarily imply consistency in higher-level bioacoustic representations. We evaluate this aspect using language-supervised AnimalCLAP~\cite{shinoda2026animalclap} and BioLingual~\cite{robinson2024biolingual}. For each encoder $E$, we compute the Fr\'{e}chet distance between the $L_2$-normalized real and generated embedding distributions separately for each species, and report their unweighted mean across species as CLAP-FAD. Because no text is provided at evaluation time, CLAP-FAD measures distribution-level fidelity in a language-aligned representation space rather than sample-wise prompt--audio or behavioral correctness. AnimalCLAP and BioLingual scores are reported separately and compared only within the same encoder.


\section{Experiments and Results}
\label{sec:experiments}

\vspace{-2mm}
\subsection{Experimental Setup}
\label{subsec:experimental-setup}

We compare \textbf{TangoFlux Zero-shot}, \textbf{TangoFlux Fine-tuned}, and \textbf{TangoFlux Event-Aware} with Event-Aware variable-length modeling. Both fine-tuned models start from pretrained TangoFlux checkpoint and share the training-sample exposure and condition-sampling strategy in Sec.~\ref{subsec:flow-matching}. The VAE and FLAN-T5 text encoder are frozen, while the Flux Transformer, text projection layer, and duration embedder are optimized for 400 epochs with batch size 64 using AdamW at a learning rate of $1\times10^{-5}$. We select the checkpoint with the lowest validation macro loss, obtained by averaging losses within each vocalization anchor and then across anchors.

At inference, each test recording is paired with one test condition, with the pairing fixed across all systems. The duration of the corresponding real test recording is used as the target duration, forming a fixed condition--duration pair for each test instance, from which one vocalization is generated using 50 generation steps and a classifier-free guidance scale of 4.5. 

\vspace{-3mm}
\subsection{Main Results}
\label{subsec:main-results}

Table~\ref{tab:main-results} summarizes the overall results. For the Real Reference, Macro Acc. and mAP are computed directly on the test set, while the distribution metrics compare the test set with an equally sized subset of the training data. Fine-tuning TangoFlux on BeVo substantially improves all evaluated metrics over its zero-shot performance, showing the effectiveness of task-specific adaptation in our experimental setting. The complete Event-Aware modeling further improves vocalization discriminability and acoustic fidelity, while the language-aligned fidelity metrics remain nearly unchanged.

\vspace{-1mm}
\noindent
\begin{minipage}{\columnwidth}
    \centering
    \setlength{\abovecaptionskip}{2pt}
    \setlength{\belowcaptionskip}{2pt}

    \captionof{table}{Species-wise performance of TangoFlux Event-Aware. Each entry reports Real Reference / Event-Aware.}
    \label{tab:species-results}

    \setlength{\tabcolsep}{5pt}
    \begin{tabular}{lccc}
        \hline
        Species
        & Macro Acc. $\uparrow$
        & FAD $\downarrow$
        & BioLi. FAD $\downarrow$ \\
        \hline
        Hyena
        & \textcolor{gray}{0.9000}/0.5500
        & \textcolor{gray}{0.1622}/1.1566
        & \textcolor{gray}{0.0227}/0.1029 \\
        Meerkat
        & \textcolor{gray}{0.7883}/0.6967
        & \textcolor{gray}{0.2478}/1.5149
        & \textcolor{gray}{0.0155}/0.0739 \\
        Marmoset
        & \textcolor{gray}{0.9717}/0.6550
        & \textcolor{gray}{0.1147}/1.8304
        & \textcolor{gray}{0.0225}/0.2129 \\
        Goat
        & \textcolor{gray}{0.8300}/0.4050
        & \textcolor{gray}{0.1561}/0.7219
        & \textcolor{gray}{0.0273}/0.0974 \\
        Zebra
        & \textcolor{gray}{0.9650}/0.6550
        & \textcolor{gray}{0.5648}/1.1341
        & \textcolor{gray}{0.0520}/0.1151 \\
        Zebra Finch
        & \textcolor{gray}{0.8560}/0.7580
        & \textcolor{gray}{0.1470}/0.9853
        & \textcolor{gray}{0.0249}/0.1364 \\
        \hline
    \end{tabular}
\end{minipage}

Table~\ref{tab:species-results} shows clear species-dependent differences in generation performance. Relative to the Real Reference, Meerkat and Zebra Finch exhibit comparatively small accuracy gaps, whereas Goat and Marmoset show larger gaps in discriminability or distribution fidelity, indicating that generation difficulty varies across species and evaluation dimensions.

\vspace{-3mm}
\subsection{Representation and Conditioning Analyses}
\label{subsec:additional-analysis}
\vspace{-1mm}

\begin{figure}[t]
    \centering
    \includegraphics[width=\linewidth]{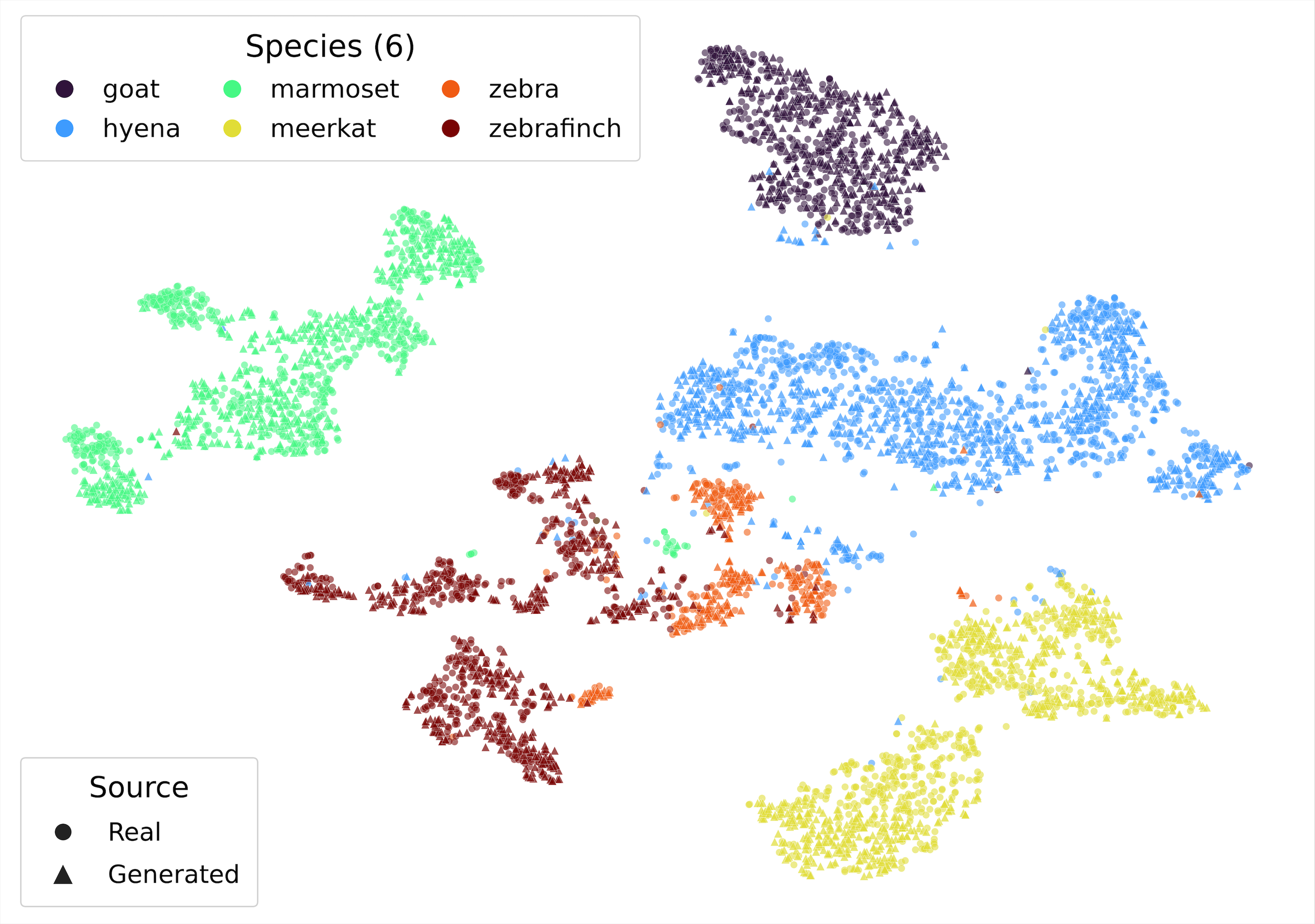}
    \vspace{-5mm}
    \caption{t-SNE of real and TangoFlux Event-Aware vocalizations in AnimalCLAP space.}
    \label{fig:animalclap_tsne}
\end{figure}

\textbf{Semantic Representation Visualization.}
To qualitatively examine representation-level fidelity, we jointly fit t-SNE on AnimalCLAP embeddings from test samples and their corresponding generated samples. As shown in Fig.~\ref{fig:animalclap_tsne}, real and generated vocalizations exhibit broadly similar species-level organization.
\par\noindent\textbf{Behavior-Conditioning Analysis.}
To assess behavioral conditioning, we compare full conditions with species-only prompts on the same test instances. We then replace each condition with one associated with a different anchor of the same species while keeping target duration and generation settings fixed; $\Delta p_{\rm tgt}$ denotes the AVEX probability change for the new target anchor.

\vspace{-1mm}
\noindent
\begin{minipage}{\columnwidth}
    \centering
    \setlength{\abovecaptionskip}{2pt}
    \setlength{\belowcaptionskip}{2pt}

    \captionof{table}{Behavior-conditioning analysis.}
    \vspace{-2mm}
    \label{tab:condition-analysis}
    \begin{tabular}{cc}
        \begin{tabular}[t]{lcc}
            \hline
            Condition & Acc. & mAP \\
            \hline
            Full & .6199 & .6551 \\
            Species-only & .2261 & .2427 \\
            \hline
        \end{tabular}
        &
        \begin{tabular}[t]{lc}
            \hline
            Behavior swap & Value \\
            \hline
            $\Delta p_{\rm tgt}$ & +.3139 \\
            Increase rate & 85.39\% \\
            \hline
        \end{tabular}
    \end{tabular}
\end{minipage}

As shown in Table~\ref{tab:condition-analysis}, full conditioning yields higher anchor discriminability, while condition swapping shifts predictions toward the new target anchor, supporting anchor-level conditional response rather than direct behavioral understanding.

\vspace{-5mm}
\section{Conclusion}
\label{sec:conclusion}
\vspace{-2mm}

We investigate behavior-conditioned animal vocalization generation through dataset construction, generative modeling, and evaluation. We construct BeVo, which links bioacoustic recordings from six species with literature-grounded behavioral descriptions through vocalization anchors, and adapt TangoFlux with Event-Aware variable-length modeling for short, duration-variable vocalizations. Fine-tuning TangoFlux on BeVo substantially improves generation performance over zero-shot generation, while the complete Event-Aware modeling further improves vocalization discriminability and acoustic fidelity, with language-aligned representation metrics remaining nearly unchanged. Conditioning analyses further show that behavioral descriptions improve target-anchor discriminability over species-only prompts and that changing the descriptions shifts predictions toward the corresponding target anchors.

\vspace{-5mm}
\section{Acknowledgement}
\label{sec:acknowledge}
\vspace{-2mm}
OpenAI ChatGPT was used solely for language refinement and manuscript editing. All scientific content, analyses, results, and conclusions were produced and verified by the authors.

\section{Compliance with Ethical Standards}
This study used publicly available animal vocalization datasets and involved no new animal experiments; no additional ethical approval was required.
\bibliographystyle{IEEEbib}
\bibliography{refs}

\end{document}